\documentclass[aps,prb,superscriptaddress,floatfix]{revtex4}

\usepackage{amsfonts}
\usepackage{amssymb}
\usepackage{graphicx}
\usepackage{texdraw}
\usepackage{color}

\newcommand{\GaMnAs}[1]{\mbox{Ga$_{1-x}$Mn$_x$As}}
\newcommand{\etal}{{\it et al.}}

\begin{document}

\title{Zero- and one-dimensional magnetic traps for quasi-particles}



\author{P. Redli\'{n}ski}
\affiliation{Department of Physics, University of Notre Dame, Notre
Dame, IN 46556}

\author{T. Wojtowicz}
\affiliation{Department of Physics, University of Notre Dame, Notre
Dame, IN 46556} \affiliation{Institute of Physics, Polish Academy of
Sciences, 02-668 Warsaw, Poland}

\author{T. G. Rappoport}
\affiliation{Instituto de F\'{\i}sica, Universidade Federal do Rio
de Janeiro, Cx. P 88528, 21945-970 Rio de Janeiro RJ, Brazil}

\author{A. Lib\'{a}l}
\affiliation{Department of Physics, University of Notre Dame, Notre
Dame, IN 46556}

\author{J. K. Furdyna}
\affiliation{Department of Physics, University of Notre Dame, Notre
Dame, IN 46556}

\author{B. Jank\'{o}}
\affiliation{Department of Physics, University of Notre Dame, Notre
Dame, IN 46556}


\begin{abstract}
We investigate the possibility of trapping quasi-particles
possessing spin degree of freedom in hybrid structures. The hybrid
system we are considering here is composed of a semi-magnetic
quantum well placed a few nanometers below a ferromagnetic
micromagnet. We are interested in two different micromagnet shapes:
cylindrical (micro-disk) and rectangular geometry. We show that in
the case of a micro-disk, the spin object is localized in all three
directions and therefore zero-dimensional states are created, and in
the case of an elongated rectangular micromagnet, the
quasi-particles can move freely in one direction, hence
one-dimensional states are formed. After calculating profiles of the
magnetic field produced by the micromagnets, we analyze in detail
the possible light absorption spectrum for different micromagnet
thicknesses, and different distances between the micromagnet and the
semimagnetic quantum well. We find that the discrete spectrum of the
localized states can be detected via spatially-resolved low
temperature optical measurement.
\end{abstract}


\maketitle

\section{Introduction}
Currently, there is an increasing interest in using the spin of
particles, in addition to their charge, as the basis for new types
of electronic devices \cite{Ohno1,Grundler,Kreuzer}. In this work we
show by theoretical calculations that the spin degree of freedom can
be utilized for achieving spatial localization - of interest for
"spintronic applications" - of charged quasi-particles (electrons,
holes or trions \cite{Kossacki, Combescot}), as well as of neutral
complexes, such as excitons \cite{Peeters1, Kossacki}.

In this paper we consider a hybrid structure consisting of a
\mbox{CdMnTe/CdMgTe} quantum well (QW) structure at a small, but
finite distance from ferromagnetic micromagnet. Due to the Zeeman
interaction, the non-homogeneous magnetic field produced by the
micromagnet acts as an effective potential that can ``trap'' spin
polarized quasi-particles in the QW. In this article we explore two
specific types of micromagnets: one with a cylindrical
\cite{Berciu1}, and one with a rectangular
\cite{Kossut1,Cywinski,Redlinski2} symmetry. In both cases the
thickness of the micromagnets is of order of a few hundreds of
nanometers and their lateral dimension is of order of microns. We
will show that in both geometries, micromagnets are very effective
in localizing quasi-particles. For the micro-disk, the
quasi-particles are localized below the center of the disk in all
three spatial directions; and for the rectangular micromagnet, the
quasi-particles are localized below the poles of the ferromagnet and
the localization occurs only in two spatial dimensions. Thus in the
latter case the quasi-particles can move quasi-freely in one
direction.

The choice of diluted magnetic semiconductor (CdMnTe) QW  instead of
the classical one (e.g. CdTe) is motivated by perspective of
achieving the more efficient spin traps, leading to the clear
localization effects. In diluted magnetic semiconductor (DMS)
materials the exchange interaction between delocalized band
electrons and localized magnetic ions (Mn$^{++}$ ions in the case at
hand), leads to a splitting between band states for different spin
components (for a review of relevant properties of DMS see
Ref.~[\onlinecite{Furdyna2}]). The so-called giant Zeeman spin
splitting  has been extensively investigated in the 80's in
\mbox{II-VI} based DMSs\cite{Furdyna1} and because of this effect
the effective \mbox{$g$-factor} for the DMS is very high. For
example, Dietl\cite{Dietl1} \etal{} reported an electron
\mbox{$g$-factor} of about 500 in a sub-Kelvin experiment in a
\mbox{CdMnSe}, which implies a value of about 2000 for the
\mbox{$g$-factor} of a hole in this material. According to our
previous estimations\cite{Redlinski2} such values for the effective
\mbox{g-factors} can in fact result in the confinement of
quasi-particles in a small lateral region inside QW. The actual
details of the optical response will depend in sensitive way on the
values of the electron and hole \mbox{g-factors}.
\begin{figure}[ht]
\includegraphics{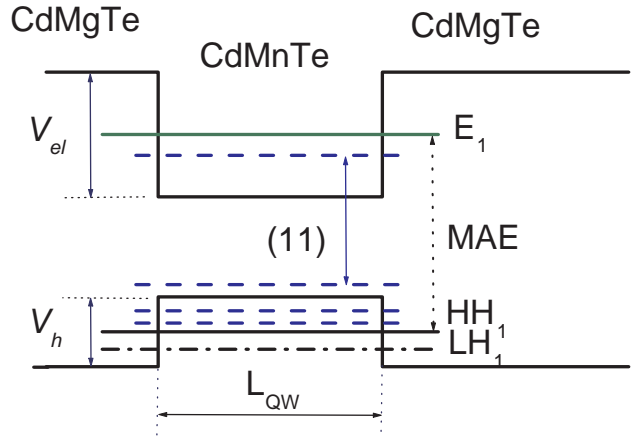}
\caption{In the absence of a micromagnet we expect the main
absorption edge (MAE) between $HH_1$ and $E_1$ energy states (full
lines). Both energy states are two-fold-degenerate without external
magnetic field. After the deposition of the micromagnet, new states
appear below the $E_1$ state and above the $HH_1$ state (dashed
lines). Each new energy state is \textit{non-degenerate}. In a QW
structure the heavy hole states and the light hole states, e.g.,
$HH_1$ and $LH_1$, are split even without external magnetic
field.}\label{figQW}
\end{figure}

Our focus on QW structures instead of normal films has both
experimental and theoretical motivations. From the experimental
point of view, we would like to avoid the complications caused by
the \mbox{metal-DMS} interface. To satisfy that criterion, the QW
should be placed at a finite distance from the micromagnet. The
local magnetic field inside the QW diminishes as the distance $d$
between the micromagnet and the QW increases. Thus, we need some
compromise for the value of $d$ in order to have a high magnetic
field in the QW, and at the same time to avoid interface
contamination effects. There are also two theoretical motivations
that require the QW to be relatively narrow. First, because of the
quantum confinement, the heavy and light hole states that are
degenerate in the bulk at the $\Gamma$ point become non-degenerate
in the QW geometry. This means that the low-energy absorption
spectrum is simplified for the QW. Second, in the narrow QW we can
assume that the local magnetic field is uniform throughout the width
of the QW which again simplifies the calculation. On the other hand,
for practical reasons the QW cannot be too narrow, since the line
width of the optical resonances increases\cite{Kuhn2} with
decreasing width of the QW ($L_{QW}$).

In Fig.~\ref{figQW} we present a schematic sketch of the energy
states in a DMS QW which can be used for discussing the presence of
ferromagnetic micromagnet on the top of the quantum structure. The
QW is grown in the (001) direction, chosen here as the $\hat{z}$
axis. Without the micromagnet, the main absorption edge is observed
between the states of the heavy hole $HH_1$ (with a pseudo spin of
$-3/2$) and the electron $E_1$ (with a spin of $-1/2$) in the
$\sigma_+$ circularly polarized light. In the absence of magnetic
field $E_1$ and $HH_1$ are two-fold degenerate with respect to the
spin, and the first optical transition in the $\sigma_-$
polarization is at the same energy as for the $\sigma_+$ transition.
After depositing the micromagnet, we expect that new states will
appear below $E_1$ and above $HH_1$. These new states are
non-degenerate, since the presence of a local magnetic field lifts
the Kramers degeneracy. It is important to note that the spin of the
states below $E_1$ remains $-1/2$, and above $HH_1$ states remains
with their pseudo-spin of $-3/2$. Our main prediction is the
following: the experimentally observable transitions between these
new states will appear in the absorption or photoluminescence
spectrum below the main absorption edge, and will obey selection
rules for $\sigma_{+}$ circularly polarized light\cite{Madelung}. We
will specifically analyze optical absorption, because it provides a
mapping of all states; but our results are equally well suited for
photoluminescence experiments if only the ground state is of
interest.

This paper is organized as follows: In the next section we present
the general theoretical approach used and the approximations made.
Then we analyze theoretically the g-factor anisotropy of the hole
states. In section \ref{resultsDysk} we show the results obtained
for the absorption coefficient of the micro-disk/QW structure as a
function of key parameters; and in section \ref{resultsRect} we
present the analysis for the rectangular ferromagnetic
micromagnet/QW hybrid. In the last section we present a detailed
discussion and their consequences.
\section{Theoretical approach}\label{theory}
The goal of this section is to derive effective Hamiltonians for
both geometries of micromagnet which can be treated numerically but
which posses all important properties of the original problem. We
begin this section by considering separately the electron in the
valence- and in the conduction band. We then discuss the Zeeman
interaction between the local magnetic field produced by micromagnet
and the quasi-particle spin. At the end of this section we discuss
the experimentally observed anisotropy of the hole g-factor ($g_h$).

We begin with the Luttinger Hamiltonian\cite{Lutt} $H_L$ of the
valence electron  in the \mbox{k-representation} (note that we are
working in the \textit{electron} representation of the valence
band), within the base of the four-component spinor
\mbox{$\Psi^\dag=(\Psi_{+3/2}^*,\Psi_{-1/2}^*,\Psi_{+1/2}^*,\Psi_{-3/2}^*)$},
\begin{equation}\label{HL}
H_L=\left(%
\begin{array}{cc}
  H^{a} & H^{b} \\
  H^{b*} & ^{T}\!H^{a} \\
\end{array}%
\right),
\end{equation}
where the matrices $H^a$ and $H^b$ (of rank two) can be written
schematically as
\begin{equation}
H^{a}=\left(%
\begin{array}{cc}
  H_h & - c  \\
  - c & H_l \\
\end{array}
\right),
\end{equation}
\begin{equation}{\label{eq:b}}
H^{b}=\left(%
\begin{array}{cc}
  -b & 0 \\
  0 & b \\
\end{array}
\right),
\end{equation}
and where
\mbox{$c=-\frac{\hbar^2}{2m_0}\sqrt{3}(\gamma_2(k_x^2-k_y^2)-2i\gamma_3k_xk_y)$},
\mbox{$b=-\frac{\hbar^2}{2m_0}2\sqrt{3}\gamma_3k_z(k_x-ik_y)$}. The
symbol $^T\!H^{a}$ denotes the $H^{a}$ matrix with interchanged
\textit{diagonal} elements. Please note that $b$ is proportional to
$k_z$ a property we will use later in the discussion. The OX, OY and
OZ axis correspond to \mbox{[1,0,0]}, \mbox{[0,1,0]} and
\mbox{[0,0,1]} crystallographic directions and the spin is quantized
along \mbox{[0,0,1]}. Using the substitution $\vec{k}\Rightarrow -i
\vec{\bigtriangledown}$, the Hamiltonian in the
\mbox{$\vec{k}$-representation} is transformed to the
\mbox{$\vec{r}$-representation}.

Now we consider the QW structure. We choose our \mbox{4-component}
spinor wave function in the following form:
\begin{equation}\label{Psi}
\Psi(\vec{r})= \left(
    \begin{array}{c}
                \Psi_{3/2}(\vec{r}) \\
                \Psi_{-1/2}(\vec{r}) \\
                \Psi_{1/2}(\vec{r}) \\
                \Psi_{-3/2}(\vec{r}) \\
    \end{array}
\right) = \left(
    \begin{array}{c}
                f_h(z)\phi_{+\frac{3}{2}}(x,y) \\
                f_l(z)\phi_{-\frac{1}{2}}(x,y) \\
                f_l(z)\phi_{+\frac{1}{2}}(x,y) \\
                f_h(z)\phi_{-\frac{3}{2}}(x,y) \\
    \end{array}
\right),
\end{equation}
where $\phi(x,y)$'s are not yet determined, and $f_h(z)$ and
$f_l(z)$ are  ground state functions of the following set of two
Schr\"{o}dinger eigen-equations,
\begin{equation}\label{Schrodinger1DHH}
\left(\frac{\hbar^2(\gamma_1-2\gamma_2)}{2m_0}\frac{d^2}{dz^2}+V_{QW}^{h}(z)\right)f_h(z)
= \epsilon_h f_h(z),
\end{equation}
\begin{equation}\label{Schrodinger1DLH}
\left(\frac{\hbar^2(\gamma_1+2\gamma_2)}{2m_0}\frac{d^2}{dz^2}+V_{QW}^{h}(z)\right)f_l(z)
= \epsilon_l f_l(z).
\end{equation}
where $\epsilon_h$, $\epsilon_l$ are the ground state energies. In
Eqs. (\ref{Schrodinger1DHH}) and (\ref{Schrodinger1DLH})
$m_0/(\gamma_1-2\gamma_2)$ and $m_0/(\gamma_1+2\gamma_2)$ are the
heavy- and light hole masses in the z-direction and $V_{QW}^h(z)$ is
a potential energy of the QW coming from discontinuity of the edge
of the valence band. We assume this potential to be rectangular (see
Fig.~\ref{figQW}),
\begin{equation}\label{Eq:StudniaH}
V_{QW}^h(z)= \left\{\begin{array}{cc}
  0 & \text{ for } |z|<L_{QW}/2 \\
  -V_h & \text{ for } |z|>L_{QW}/2\;. \\
\end{array}\right.\nonumber
\end{equation}
The functions $f_h(z)$ and $f_l(z)$ which satisfy Eqs.
(\ref{Schrodinger1DHH}) and (\ref{Schrodinger1DLH}) are real-valued
even functions fulfilling the relation $\int f_h^*(z)\,p_z\, f_l(z)
dz = 0$. These properties of $f_h$ and $f_l$ can be used as follows:
In the subspace of Hilbert space spanned by our trial wave function,
$H_L$ Eq.~(\ref{HL}) is written as a new 4x4 matrix $\tilde{H}_L$,
\begin{equation}\label{Eq:tildeHL}
\tilde{H}_L=\left(%
\begin{array}{cc}
  \tilde{H}^{a} & \tilde{H}^{b} \\
  \tilde{H}^{b} & ^{T}\!\tilde{H}^{a} \\
\end{array}%
\right),
\end{equation}
where
\begin{equation}\label{DefIhl}
\tilde{H}^{a}=\left(%
\begin{array}{cc}
  H_h & - c\cdot I_{hl} \\
  -c^\cdot I^*_{hl} & H_l \\
\end{array}
\right),
\end{equation}
and
\begin{equation}
\tilde{H}^{b}=\left(%
\begin{array}{cc}
  0 & 0 \\
  0 & 0 \\
\end{array}
\right).
\end{equation}
Symbol $I_{hl}$ in Eq.~(\ref{DefIhl}) is the overlap integral,
defined by $I_{hl}\equiv\int f_h^*(z)f_l(z)dz$. Furthermore, in the
\mbox{$\vec{r}$-representation} the two quantities $H_h$ and $H_l$
in the matrix $\tilde{H}^{a}$ take the following form:
\begin{eqnarray}
H_h & = &
\frac{\hbar^2}{2m_0}(\gamma_1+\gamma_2)\left(\frac{d^2}{dx^2}+\frac{d^2}{dy^2}\right)+\epsilon_h,\\\label{Eq:HH}
H_l & = &
\frac{\hbar^2}{2m_0}(\gamma_1-\gamma_2)\left(\frac{d^2}{dx^2}+\frac{d^2}{dy^2}\right)+\epsilon_l,\\\label{Eq:LH}
-c & = -&
\frac{\hbar^2}{2m_0}\sqrt{3}\left(\gamma_2(\frac{d^2}{dx^2}-\frac{d^2}{dy^2})-2i\gamma_3\frac{d}{dx}\frac{d}{dy}\right).\nonumber
\end{eqnarray}
After making these approximations the wave function $\tilde{\Psi}$
satisfying $\tilde{H}_L \tilde{\Psi}=E_h\tilde{\Psi}$ has only two
non-vanishing components,
\begin{equation}
\tilde{\Psi}(\vec{r})= \left(
    \begin{array}{c}
                \phi_{+\frac{3}{2}}(x,y) \\
                \phi_{-\frac{1}{2}}(x,y) \\
                0 \\
                0 \\
    \end{array}
\right),
\end{equation}
and
\begin{equation}
\tilde{\Psi}(\vec{r})= \left(
    \begin{array}{c}
                0 \\
                0 \\
                \phi_{+\frac{1}{2}}(x,y) \\
                \phi_{-\frac{3}{2}}(x,y) \\
    \end{array}
\right).
\end{equation}
The Zeeman interaction due to the local magnetic field produced by
the micromagnet mixes both states and lifts their degeneracies.

Now we consider the conduction band. We assume that the dispersion
relation of the conduction electron is parabolic, its Hamiltonian
$H_e$ taking the form
\begin{equation}\label{Eq:HamEl}
H_e=E_G-\frac{\hbar^2}{2m_0}\gamma_{el}(\frac{d^2}{dx^2}+\frac{d^2}{dy^2}+\frac{d^2}{dz^2})
+ V_{QW}^{el}(z),\nonumber
\end{equation}
where $E_G$ is the energy gap of a QW, $m_0/\gamma_{el}$ is the
electron effective mass, and $V_{QW}^{el}$ is the potential energy
coming from the discontinuity of the conduction band (see
Fig.~\ref{figQW}):
\begin{equation}\label{Eq:StudniaEL}
V_{QW}^{el}(z)= \left\{\begin{array}{cc}
  0 & \text{ for } |z|<L_{QW}/2 \\
  V_{el} & \text{ for } |z|>L_{QW}/2
\end{array}\right). \nonumber
\end{equation}
For the electron we assume the following factorized trial wave
function
\begin{equation}\label{Eq:ffel}
\Psi_{el}=\left(%
\begin{array}{c}
  f_{el}(z)\phi_{+}(x,y) \\
  f_{el}(z)\phi_{-}(x,y)
\end{array}%
\right),
\end{equation}
where the function $f_{el}(z)$ is the solution of the
one-dimensional Schr\"{o}dinger equation
\begin{equation}\label{Schrodinger1DEL}
\left(-\frac{\hbar^2\gamma_{el}}{2m_0}\frac{d^2}{dz^2}+V_{QW}^{el}(z)\right)f_{el}(z)
= \epsilon_{el} f_{el}(z).
\end{equation}
Both functions $\phi_{\pm}$ fulfill the following eigen-equation
\begin{equation}
  \tilde{H}_e \phi_{\pm}(x,y) \equiv \left(-\frac{\hbar^2\gamma_{el}}{2m_0}(\frac{d^2}{dx^2}+\frac{d^2}{dy^2})+\epsilon_{el}\right)\phi_{\pm}(x,y)
  = E_e \phi_{\pm}(x,y).
\end{equation}

The Zeeman Hamiltonian used for both the conduction and the valence
electrons can be written as
\begin{equation}\label{Eq:Zeeman}
H_Z(\vec{r})=\mu_B\,\vec{s}\,\hat{g}_{eff}\,\vec{B}(\vec{r}),
\end{equation}
where $\hat{g}_{eff}$ is a tensor, $\mu_B$ is the Bohr magneton,
$\vec{B}(\vec{r})$ is the local magnetic field produced by the
micromagnet, and $\vec{s}$ is the \mbox{spin-$\frac{1}{2}$}
operator. For the conduction band \mbox{$\hat{g}_{eff}=g_e
\hat{1}\equiv diag(g_e,g_e,g_e)$} and for valence electron
\mbox{$\hat{g}_{eff}=g_h \hat{1}\equiv diag(g_h,g_h,g_h)$}, where
$\hat{1}$ is the 3x3 identity matrix. For the conduction electron,
the \mbox{spin-$\frac{1}{2}$} operator in the basis of the Bloch
states is a \mbox{2x2-matrix} proportional to the Pauli matrices.
For the valence electron the spin operator is a \mbox{4x4-matrix},
as shown in Ref.~[\onlinecite{Abolfath}]. In addition to the
previous approximations we assume that the QW is narrow, and we
rewrite Eq.~(\ref{Eq:Zeeman}) as
\begin{equation}\label{Eq:Zeeman2}
\tilde{H}_Z(x,y)=\mu_B\,\vec{s}\,\hat{g}_{eff}\,\vec{B}(x,y,z=d),
\end{equation}
where $d$ is the distance between the QW and the micromagnet.

Original Hamiltonians are transformed into the 2D effective
Hamiltonians because of the form of the trial wave function we used.
The total effective Hamiltonians of the valence ($\mathcal{H}_{h}$)
and of the conduction electron ($\mathcal{H}_{e}$) then become
\begin{eqnarray}
\mathcal{H}_{h}(x,y) & = & \tilde{H}_L(x,y) + \tilde{H}_Z(x,y), \label{Eq:totalH1}\\
\mathcal{H}_{e}(x,y) & = & \tilde{H}_{e}(x,y) + \tilde{H}_Z(x,y).
\label{Eq:totalH2} \label{Eq:totalE}
\end{eqnarray}
The procedure of the calculation is as follows: first, the magnetic
field produced by the micromagnet is calculated by solving the
magneto-static Maxwell equations\cite{Jackson}, either by using a
magnetization distribution determined from micromagnetic
simulations\cite{Berciu1} in the case of the disk or by assuming
that the micromagnet can be represented by a uniformly magnetized
domain in the case of rectangular micromagnet. Then, the electronic
spectrum of the conduction band and the valence band is calculated
by approximating the Schr\"{o}dinger eigen-equations with finite
difference algebraic equations. In the case of the micro-disk, the
finite difference algebraic equations are generated on a
two-dimensional grid and Eqs. (\ref{Eq:totalH1}), (\ref{Eq:totalH2})
are solved directly. In the case of the rectangular micromagnet -
because of its large elongation in the y-direction ($D_y$=2$\mu$m) -
we can make the ansatz that the quasi-particle is free to move in
this direction so the plane wave form of the wave function is
assumed in y-direction. Thus, in the case of a rectangular
micromagnet the finite difference equations are generated only on a
one-dimensional grid.

Given the calculated eigen-values and eigen-functions, we use the
Fermi's Golden rule to obtain the absorption
coefficient\cite{Madelung}
\begin{equation}\label{abscoef}
\alpha_{\pm}(\omega)\simeq\frac{1}{\omega}\sum_{i,j}|<e_j|p_x \pm i
p_y|h_i>|^2\delta(\hbar\omega-(E_{e_i}-E_{h_i})),
\end{equation}
where $\{|e_i>, E_{e_i}\}$ and $\{|h_j>, E_{h_j}\}$ are the
eigen-solutions of the conduction and valence band Hamiltonians,
$\mathcal{H}_e$ and $\mathcal{H}_h$, respectively. The Coulomb
interaction between the electrons and the holes leads to the
creation of long-lived excitons, which in turn generate sharp
individual optical lines. We assumed that the exciton states are
formed and that the MAE corresponds to the 1S exciton transition in
the QW without the micromagnet. In the following figures we have
taken the zero energy at the 1S exciton main absorption peak. To
create the final spectrum, each optical line which we calculate is
broadened by a gaussian function with a linewidth of 1~meV. This is
a reasonable approximation for typical experimental resolution.

\subsection{Anisotropy of the g-factor of the hole}\label{anisotropy}
The Zeeman splitting of the valence band edge depends on the
direction of the magnetic field with respect to the growth direction
of the quantum well\cite{Kuhn1}. In order to demonstrate this we
consider two configurations: a constant magnetic field parallel to
the plane of the quantum well, \mbox{$\vec{B}=B\vec{e}_x$}; and a
constant magnetic field perpendicular to this plane,
\mbox{$\vec{B}=B\vec{e}_z$}.

Starting with $\vec{B}=B\vec{e}_z$, we find it convenient to use a
different basis than that used in the previous part of the paper:
the new basis vectors are
\mbox{$\Psi^{\dag}=(\Psi_{+3/2}^*,\Psi_{-3/2}^*,\Psi_{+1/2}^*,\Psi_{-1/2}^*)$}.
In this new basis, the Luttinger Hamiltonian, Eq.~(\ref{HL}) or
Eq.~(\ref{Eq:tildeHL}), for the edge of the valence band in the QW
is a diagonal matrix
\begin{widetext}
\begin{equation}\label{H1}
H=\left(%
\begin{array}{cccc}
  +\frac{1}{2}g_h \mu_B B & 0 & 0 & 0 \\
  0 & -\frac{1}{2}g_h \mu_B B & 0 & 0 \\
  0 & 0 & -\Delta_{lh} +\frac{1}{6}g_h \mu_B B & 0 \\
  0 & 0 & 0 & -\Delta_{lh} -\frac{1}{6}g_h \mu_B B\\
\end{array}%
\right),
\end{equation}
\end{widetext}
where the energy splitting $\Delta_{lh}=|E_l-E_h|$ is caused by the
quantum well confinement. This splitting is also present in
structures under strain and it can be either positive or negative as
shown in Ref.~\onlinecite{Abolfath}. Looking at the Hamiltonian $H$,
Eq.~(\ref{H1}), we can see that the energy of the heavy hole
$(+\frac{3}{2}, -\frac{3}{2})$ splits by $|g_h \mu_B B|$, whereas
the energy of the light hole $(+\frac{1}{2}, -\frac{1}{2})$ splits
by $|\frac{1}{3}g_h \mu_B B|$, an amount smaller by a factor of 3.

\begin{figure}[ht]
\includegraphics{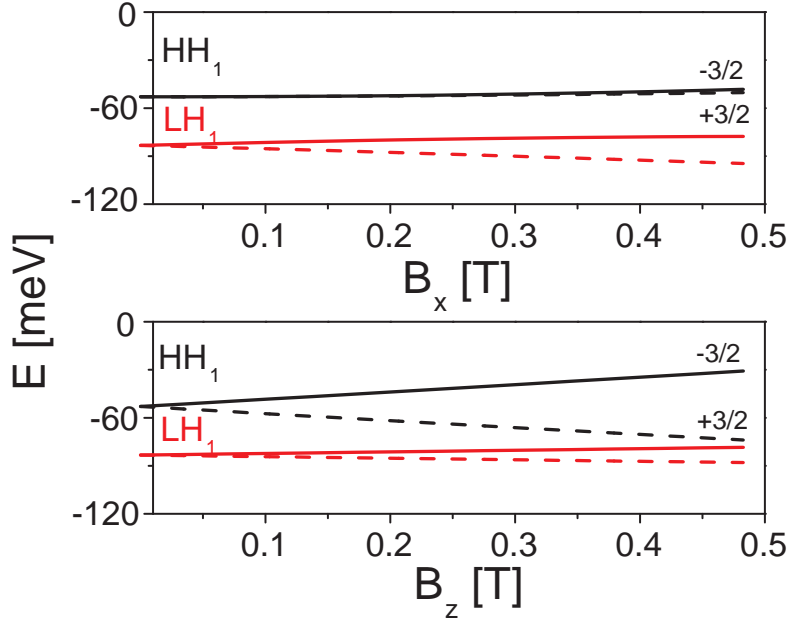}
\caption{Top panel: splitting of the heavy hole ($HH_1$) and light
hole ($LH_1$) edges at the $\Gamma$ point in a QW vs. $B_x$ (we use
the electron representation). $B_y$ and $B_z$ are set to zero.
Bottom panel: splitting of the heavy hole and light hole edges in a
QW vs. $B_z$ ($B_x$=0 and $B_y$=0). Solid and dashed lines represent
states with opposite spin. Heavy hole states $\mp 3/2$ \textit{do
not split} when the magnetic field is applied in the plane of the QW
as seen on the upper panel.}\label{figAnisotropy}
\end{figure}

In order to analyze the second geometry ($\vec{B}=B\vec{e}_x$), we
write the Hamiltonian for the valence band edge in a QW as follows:
\begin{widetext}
\begin{equation}\label{HedgeBx}
H=\left(%
\begin{array}{cccc}
  \mathbf{0} & \mathbf{0} & \frac{1}{2\sqrt{3}}g_h \mu_B B & 0 \\
  \mathbf{0} &  \mathbf{0} & 0 & \frac{1}{2\sqrt{3}}g_h \mu_B B \\
  \frac{1}{2\sqrt{3}}g_h \mu_B B & 0 & \mathbf{-\Delta_{lh} } & \mathbf{\frac{1}{3}g_h \mu_B B} \\
  0 & \frac{1}{2\sqrt{3}}g_h \mu_B B & \mathbf{\frac{1}{3}g_h \mu_B B} & \mathbf{-\Delta_{lh}}
\end{array}%
\right).
\end{equation}
\end{widetext}
If we omit the mixing between the heavy and light holes (i.e., if we
retain only the bold elements in Eq.~(\ref{HedgeBx})) then the heavy
holes do not split, and the light holes split by $|\frac{2}{3}g_h
\mu_B B|$. We note that the light hole splitting is now two times
larger than in the case of the perpendicular magnetic field,
$\vec{B}=B\vec{e}_z$, where it was $|\frac{1}{3}g_h \mu_B B|$.
Solving eigenvalue problem with Hamiltonian $H$,
Eq.~(\ref{HedgeBx}), we obtain four solutions as a function of the
external magnetic field (see Fig.~\ref{figAnisotropy}):
\begin{widetext}
\begin{eqnarray}
E_h^{\mp} & = & - \frac{\Delta_{lh}}{2} \mp \frac{g_h \mu_B B}{6} +
\frac{1}{6}\sqrt{9\Delta_{lh}^2 \pm 6\Delta_{lh}g_h \mu_B B +
4(g_h\mu_B B)^2},\label{ExactSolutionsBx1}\\
E_l^{\mp} & = & - \frac{\Delta_{lh}}{2} \mp \frac{g_h \mu_B B}{6} -
\frac{1}{6}\sqrt{9\Delta_{lh}^2 \pm 6\Delta_{lh}g_h \mu_B B +
4(g_h\mu_B B)^2}.\label{ExactSolutionsBx2}
\end{eqnarray}
\end{widetext}
When the QW splitting $\Delta_{lh}>>|g_h \mu_B B|$ then up to terms
quadratic in the external field $B$, Eqs.~(\ref{ExactSolutionsBx1})
and (\ref{ExactSolutionsBx2}) can be further simplified to
\begin{eqnarray}
E_h^{\mp}&\approx&\frac{1}{12}\frac{g_h \mu_B B}{\Delta_{lh}}g_h \mu_B B,\label{hhsplit}\\
E_l^{\mp}&\approx&-\Delta_{lh}\mp\frac{g_h \mu_B
B}{3}-\frac{1}{12}\frac{g_h \mu_B B}{\Delta_{lh}}g_h \mu_B
B\label{lhsplit}.
\end{eqnarray}
Because the heavy hole does not split, we are left with 3 solutions
out of the 4. In this approximation the energy level of the heavy
hole $E_h^{\mp}$ is twofold degenerate and the light hole splits by
an amount proportional to the magnetic field ($\frac{2}{3}|\mu_B g_h
B|$), as seen in Eq.~(\ref{lhsplit}). These different behaviors can
be compared in Fig.~\ref{figAnisotropy} where we plotted exact
values of $E_h^{\pm}$ and $E_l^{\mp}$ using Eqs.
(\ref{ExactSolutionsBx1}) and (\ref{ExactSolutionsBx2}).

In other words, we can say that in a QW, the \mbox{$g_h$-factor} of
the first heavy hole states $HH_1$ is highly anisotropic.
Consequently, in the Zeeman term given by Eqs.~(\ref{Eq:Zeeman}) or
(\ref{Eq:Zeeman2}), the tensor \mbox{$\hat{g}_h=diag(g_h,g_h,g_h)$}
can be approximated by \mbox{$\hat{g}_h\approx diag(0,0,g_h)$}.
Indeed, only the component of the field parallel to the growth
direction of the QW splits the $HH_1$ states. The anisotropy that
appears naturally using the Luttinger Hamiltonian in the QW is
observed experimentally\cite{Kuhn1}. We compared the results of the
optical response for both isotropic and anisotropic
\mbox{g-factors}, and found that absorption spectra are similar,
especially in the low-energy region of the spectrum. This result is
presented in the next paragraph.
\section{Results and Discussion}\label{results}
For the calculations involving both cylindrical and rectangular
micromagnets, the following set of Luttinger parameters and electron
mass was chosen for both the QW and the barriers:
\mbox{$\gamma_1$=4.14}, \mbox{$\gamma_2$=1.09},
\mbox{$\gamma_3$=1.62}, \mbox{$m_e=m_0/\gamma_e=0.096\,m_0$}. From
Eq.~(\ref{Eq:HH}), the heavy hole effective mass in the plane is
\mbox{$m_0/(\gamma_1+\gamma_2)$=0.19$\,m_0$} and light hole
effective mass in the plane
\mbox{$m_0/(\gamma_1-\gamma_2)$=0.33$\,m_0$}. A total discontinuity
of bands \mbox{$V_T$=500~meV}, and valence band offset
\mbox{vbo=0.4} is assumed, which corresponds to a discontinuity in
the valence band of \mbox{$V_h$=$V_T$\,vbo=200~meV}, and a
discontinuity in the conduction band of
\mbox{$V_{el}=V_T\cdot(1-vbo)$=300~meV} (see Fig.~\ref{figQW}). We
also choose a quantum well width of $L_{QW}$=20~\AA, for which the
splitting between the heavy hole $HH_1$ and the light hole $LH_1$
energy states is $\Delta_{lh}\approx$50~meV. Note that there is only
one bound heavy hole state and only one bound light hole state for
these parameters.
\subsection{Cylindrical micromagnet: a zero-dimensional trap}\label{resultsDysk}
We investigate a Fe micro-disk (with a diameter of $R$=1~$\mu$m, a
thickness $D_z$=50~nm, and $\mu_0 M_s$=2.2~T) in the vortex state
\cite{Berciu1,Ha}. In this state, due to the competition between the
exchange energy and the demagnetization energy, the magnetization
lies in the plane of the micro-disk except near the center, where
the local magnetization points out of the plane (to reduce the
exchange energy), and forms a magnetic vortex.
\begin{figure}[ht]
\includegraphics{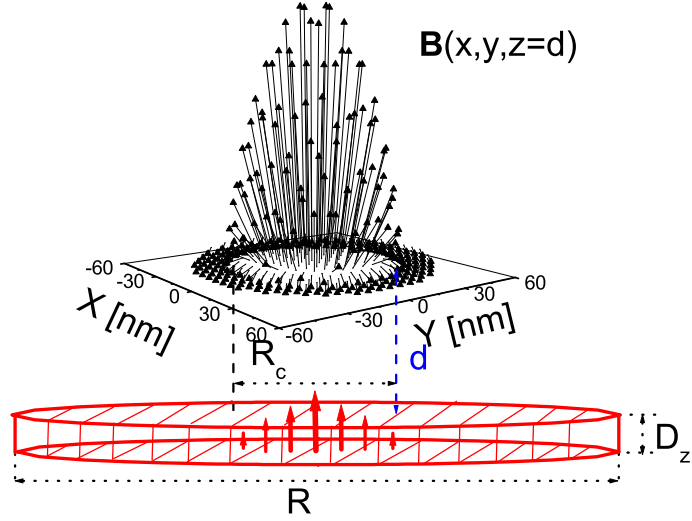}
\caption{Three-dimensional image of a magnetic field $B(x,y,z=d)$ on
the XY surface at a distance $d$ above Fe micro-disk. It is assumed
that micromagnet is in a vortex state. The diameter of the region
where the z-component of the magnetization $M_z$ is non-zero (the
core region) is only of the order of $R_c=60$~nm, whereas the
diameter of the micro-disk $R$ is 1 $\mu$m, and the thickness of the
disk $D_z$ is 50~nm. $M_z$ depicted by vertical arrows in the core
region is very well approximated by a parabolic
profile.}\label{FigB3Ddisk}
\end{figure}
The diameter of the core ($R_c$) to which the non-zero perpendicular
magnetization is confined extends over only about 60~nm, as
previously calculated \cite{Berciu1}. It is important to mention
that only the z-component $M_z$ of the total magnetization $\vec{M}$
produces the "spike" in the magnetic field $\vec{B}$ which traps the
quasi-particles.

Having magnetization profile of the vortex\cite{Berciu1, Ha} and
using magneto-static Maxwell equation magnetic field $\vec{B}$ is
calculated. In Fig.~\ref{FigB3Ddisk} we show the distribution of the
magnetic field $\vec{B}(x,y,d)$ in the XY plane at a distance $d$
above the micromagnet, where $d$ is the separation between the
micromagnet and the QW. In the magnetostatic picture, the magnetic
field $\vec{B}$ is produced by two magnetic charges (of diameter
$R_c$) on two surfaces of the micromagnet. These magnetic charges
are separated by a thickness of the micromagnet, $D_z$. At a
distance $d$=10~nm above the micromagnet, the maximum value of the
magnetic field is \mbox{$|\vec{B}|_{max}$=0.46~T}. The field is
nonzero over a distance of \mbox{60 to 80~nm} from the center of the
disk, as shown in Fig.~\ref{FigB3Ddisk}. Such a strongly localized
magnetic field in both x and y directions, together with the QW
confinement results in a quasi-particle localization in all three
directions. The particle is localized below the center of the
micro-disk. We will refer to this system as a zero-dimensional trap.

\begin{figure}[ht]
\includegraphics{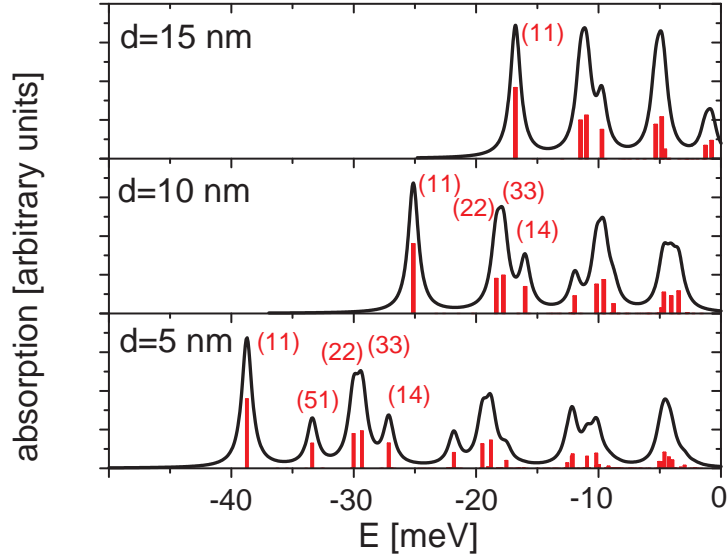}
\caption{Absorption coefficient for three distances $d$ between the
micro-disk and the QW with $D_z$=50~nm. The number ($n_cn_v$)
corresponds to transitions between the $n^{th}_c$ electron state and
the $n^{th}_v$ hole state.}\label{FigDiskAbsVsd}
\end{figure}

In Fig.~\ref{FigDiskAbsVsd} we show the absorption coefficient for
three distances $d$ between the micro-disk and the QW: $d$=5~nm,
10~nm and 15~nm.
The energy of the photon is measured relative to the energy of the
main absorption peak in the QW in the absence of a micromagnet (see
Fig.~\ref{figQW}). Vertical bars represent the optical oscillator
strength of the transitions, and the numbers ($n_cn_v$) show that
the corresponding line is a transition between the $n^{th}_v$ hole
state and the $n^{th}_c$ electron state. Eigenenergies of the
conduction electron fulfil $E_1 < E_2 < ... < E_{n_c} < ... $ and
$E_1$ is the ground state of the conduction band electron.
Eigenenergies of the valence electron fulfil $E_1 > E_2 > ... >
E_{n_v} > ... $ and $E_1$ is the ground state of the valence band
electron.

The absorption line was obtained after broadening the
\mbox{$\delta$-distribution} of each transition,
Eq.~(\ref{abscoef}), with a Gaussian function (the line width of
each resonance is $w$=1~meV). As expected, the peaks shift to lower
energies with decreasing $d$, as the maximum value of $B$, and thus
the effective 'potential', are larger for smaller $d$. At $d$=10~nm,
the shift between the (11) transition and the main absorption peak
is around 25~meV (we will call this quantity the \textit{binding
energy}). As is seen in Fig.~\ref{FigDiskAbsVsd}, the first
transition (11) is the most intense, but we can see that
non-diagonal transitions (14),... also have a relatively large
spectral weight. We also see that for a set of parameters we used in
calculation the two closely lying lines (22) and (33) merge into a
single large peak in the absorption spectrum.

\begin{figure}[ht]
\includegraphics{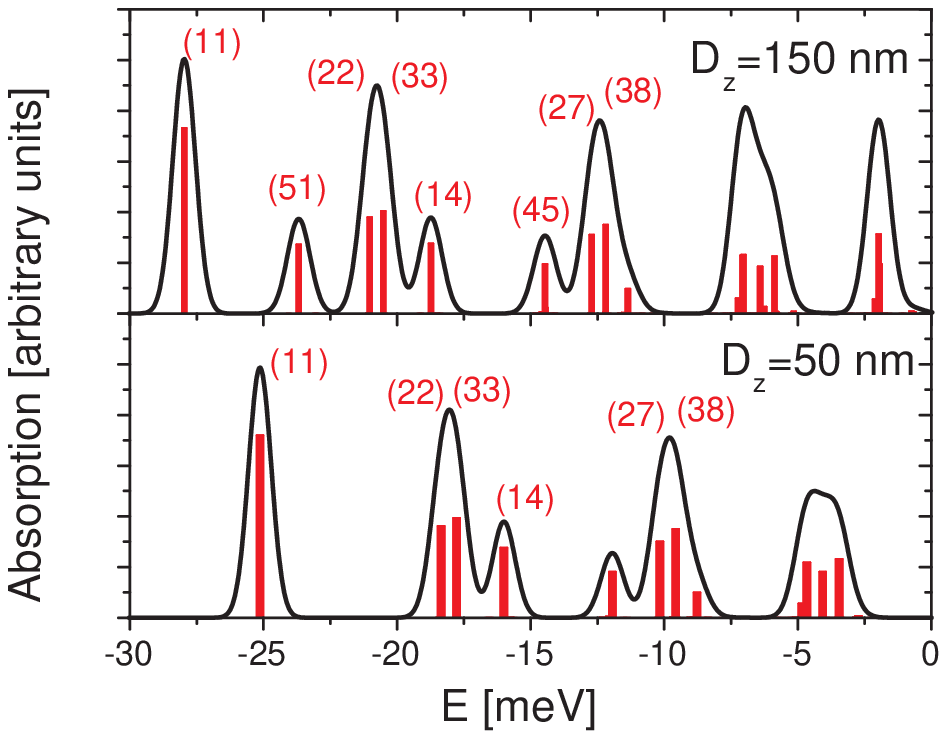}
\caption{Absorption coefficient for two thicknesses $D_z$ of the
ferromagnetic micro-disk. In magnetostatic language, $D_z$ is the
distance between magnetic charges. The separation between the
micro-disk and the QW was set to $d$=10~nm.}\label{FigDiskAbsVsDz}
\end{figure}

In Fig.~\ref{FigDiskAbsVsDz} we show the absorption coefficient for
two different thicknesses $D_z$ of the micro-disk and a value of $d$
fixed at 10~nm.
When the thickness $D_z$ increases from 50~nm to 150~nm, the binding
energy increases from 25~meV to 28~meV. Transition (51), not seen
for $d$=10 nm in Fig.~\ref{FigDiskAbsVsd}, now becomes visible at
$D_z$=150~nm. On the other hand, increasing $D_z$ to larger values
does not change the absorption substantially, because the separation
between magnetic charges ($D_z$) begins to exceed the dimension of
the magnetic charge ($R_c$).

To get further insight into the optical transitions we follow the
prescription defined in Ref.~\onlinecite{Berciu1} and write the
conduction electron two component spinor as
\begin{equation}\label{psimkel}
\psi_{m,k}^c=f_{el}(z)e^{im\phi}\left(%
\begin{array}{c}
  e^{-i\phi}g^{\uparrow}_{m,k}(\rho) \\
  g^{\downarrow}_{m,k}(\rho) \\
\end{array}%
\right),
\end{equation}
where $m$=0, $\pm$1, $\pm$2,... is the angular momentum quantum
number, $k$=0, 1, 2... is the radial quantum number and we used
cylindrical coordinates ($\rho$, $\phi$, z). This form of the
angular dependence of the wave function is exact for electrons
described by spherically symmetric bands with quadratic dispersion.
Assuming that the valence band can also be approximated by a
quadratic dispersion, the hole four-component spinor has the
following form
\begin{equation}\label{psimkh}
\psi_{m,k}^v=e^{im\phi}\left(%
\begin{array}{c}
  f_h(z)e^{-3i\phi}g^{+3/2}_{m,k}(\rho) \\
  f_l(z)e^{-2i\phi}g^{+1/2}_{m,k}(\rho) \\
  f_l(z)e^{-i\phi}g^{-1/2}_{m,k}(\rho) \\
  f_h(z)g^{-3/2}_{m,k}(\rho) \\
\end{array}%
\right).
\end{equation}
The dominant part of the electron and the hole wave function
calculated in our approach is of the form given in
Eqs.~(\ref{psimkel}) and (\ref{psimkh}).

In Ref.~\onlinecite{Berciu1} the intra-band optical transitions of
the conduction band were studied. Selection rules for this type of
the transitions imply that the angular momentum quantum number of
the envelope wave function $m$ must be changed by $\pm1$ in
$\sigma_{\pm}$ polarizations ($\Delta m = \pm1$). On the other hand
for inter-band transitions $\Delta m = 0$ because the initial and
the final Bloch states are of different symmetry: the initial state
is of P symmetry and the final state is of S symmetry. In order to
calculate the $\sigma_{+}$ absorption coefficient using
Eq.~(\ref{abscoef}) we have to calculate two integrals:
$<\Psi_{-3/2}|\Psi_{\downarrow}>$ and
$<\Psi_{-1/2}|\Psi_{\uparrow}>$. Using Eqs.~(\ref{psimkel}) and
(\ref{psimkh}) as an approximation for the true wave function we
have
\begin{eqnarray}
<\Psi_{-3/2}|\Psi_{\downarrow}> & \approx & \int
e^{-im\phi}(f_h(z)g^{-3/2}_{m,k}(\rho))^{*}f_{el}(z)e^{im^{'}\phi}g^{\downarrow}_{m^{'},k^{'}}(\rho)\,
d\phi\, rdr\, dz \sim \delta_{m,m^{'}}, \nonumber\\
<\Psi_{-1/2}|\Psi_{\uparrow}> & \approx & \int
e^{-im\phi}(f_h(z)g^{-1/2}_{m,k}(\rho))^*f_{el}(z)e^{im^{'}\phi}g^{\uparrow}_{m^{'},k^{'}}(\rho)\,
d\phi\, rdr\, dz \sim \delta_{m,m^{'}}, \nonumber
\end{eqnarray}
from where the conservation of the $m$ number is immediately
obtained: $m=m^{'}$.

\begin{table}
\caption{\label{table1} Mapping of the quantum number $n$ that was
used in Fig.~\ref{FigDiskAbsVsd} and Fig.~\ref{FigDiskAbsVsDz} to
described hole states on the (m,k) pairs. Only four states, n=1..4,
which fulfill $E_1>E_2>E_3>E_4$ are mapped, and $E_1$ is the ground
state of the electron in the valence band.}
\begin{ruledtabular}
\begin{tabular}{ccc}
$n_v$  & m & k \\
\hline
1   &   0  &  0\\
2   &   -1  &  0 \\
3   &   +1  &  0 \\
4   &   0  &   1
\end{tabular}
\end{ruledtabular}
\end{table}

In Tab.~\ref{table1} we show the mapping between the notation of the
hole states used previously $n_v$ and the ($m$, $k$) notation. A
similar table can be drawn for the electron states. As an example,
the non-diagonal optical transition (14) shown both in
Fig.~\ref{FigDiskAbsVsd} or Fig.~\ref{FigDiskAbsVsDz} is relatively
strong because both the first electron state (1) and the fourth hole
state (4) have the same $m$ quantum number, i.e. $m=0$.

\subsection{Rectangular micromagnet: a one-dimensional trap}\label{resultsRect}
In this section we consider a rectangular, flat Fe micromagnet in
the single domain state\cite{Jackson}, with magnetization pointing
in the \mbox{x-direction} \cite{Cywinski,Crowell}, see also
Ref.~\onlinecite{Redlinski2}. The single-domain state of the
micromagnet with the mentioned size was investigated by
micro-magnetic simulation using the OOMMF package\cite{oommf}.
Without an external magnetic field the sample does not remain as an
ordered single-domain, but rather goes into a multi-domain
structure. The simulation shows however, that after magnetizing the
sample with a field of 1~T and reducing the field close to 0~T, a
value of 0.2~T is sufficient to restore a state that - for our
purposes - is sufficiently close to a single domain. Because of the
magnetic anisotropy of the $g_h$, this additional field is
unimportant for electrons in the valence band, but it does have a
slight effect on the conduction electron spectrum.

\begin{figure}[ht]
\includegraphics{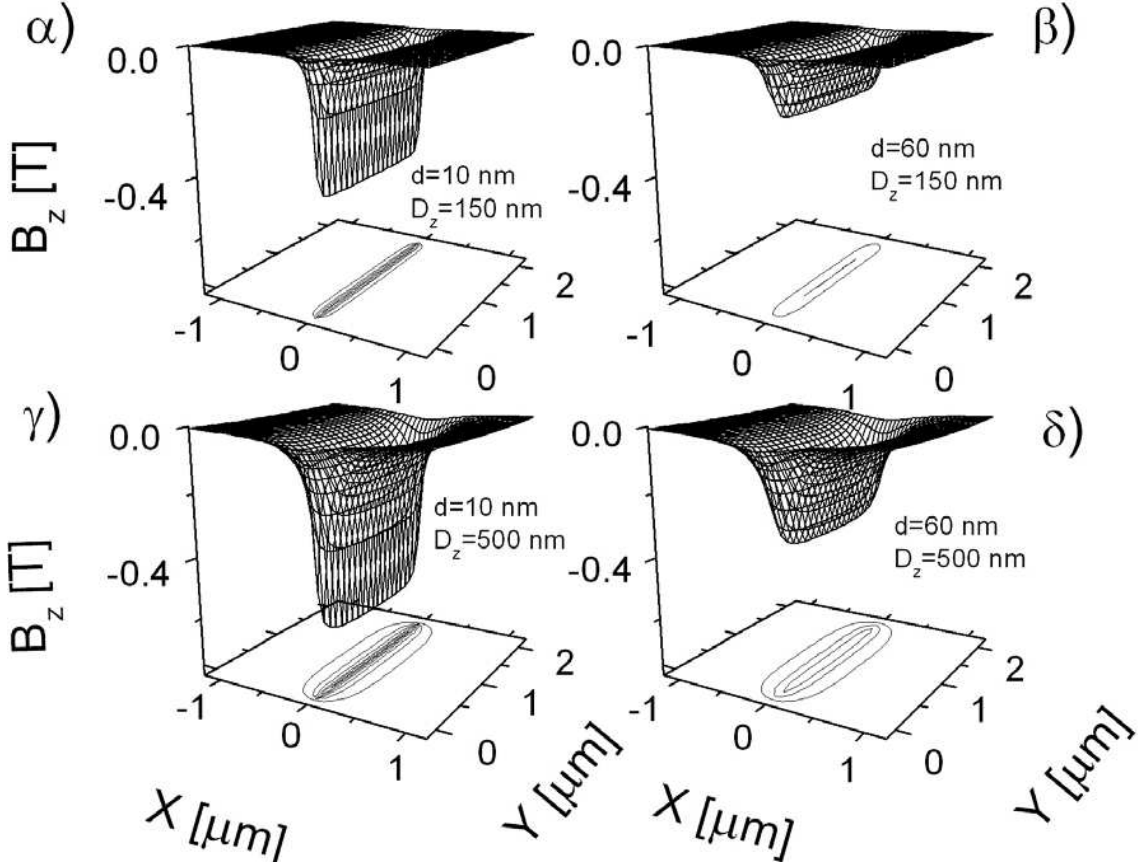}
\caption{Z-component of the magnetic field $\vec{B}$ produced by a
rectangular micromagnet at a distance $d$ below its narrow edge. We
show results for two distances $d$=10~nm, $d$=60~nm and for two
micromagnet thicknesses, $D_z$=150~nm and $D_z$=500~nm. Contour
plots on the XY planes indicate localization and 'strength' of the
z-component of the magnetic field.}\label{FigBz4rect}
\end{figure}

\begin{figure}[ht]
\includegraphics{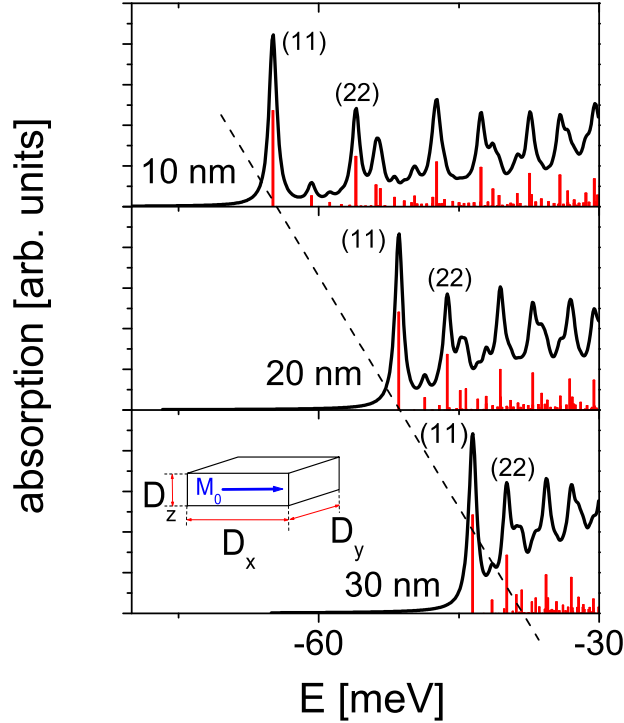}
\caption{Absorption coefficient for three distances $d$=10, 20 and
30~nm (from top to bottom) between the rectangular micromagnet
($D_z$=150~nm, $D_x$=6$\mu$m and $D_y$=2$\mu$m) and the QW. The
binding energy as well as the separation between the peaks decreases
almost linearly with increasing $d$. In the inset we show the
arrangement of the rectangular micromagnet in a single domain state.
Magnetization is pointing in the x-direction.}\label{FigRectAbsVsd}
\end{figure}

In Fig.~\ref{FigBz4rect} we present the z-component of the magnetic
field below one of the two poles of the micromagnet
($D_x$=6~$\mu$m$\times$$D_y$=2~$\mu$m$\times D_z$, see inset in
Fig.~\ref{FigRectAbsVsd}). The thickness $D_z$ is a parameter in our
simulations, and we vary this value in the range from 150 to 500~nm.
For large $D_x$, the local magnetic field $\vec{B}(\vec{r})$ can be
thought to be a sum of two fields (one of them is shown in
Fig.~\ref{FigBz4rect}) produced by the magnetic charges localized at
the two magnetic poles of the micromagnet. The magnetic field $B_z$
on the second pole has an opposite direction to this field, and thus
'attracts' quasi-particles with the opposite spin. The top two plots
($\alpha$ and $\beta$) in Fig.~\ref{FigBz4rect} are for two
different distances $d$=10~nm, $d$=60~nm keeping a constant
thickness of $D_z$=150~nm. The two bottom plots ($\gamma$ and
$\delta$) are for the same distances, using $D_z$=500~nm. When the
distance $d$ is increased while keeping the thickness $D_z$ constant
(as we go from $\alpha$ to $\beta$ or from $\gamma$ to $\delta$),
the magnetic field $B_z$ is seen to decrease, as expected. In
contrast, when the thickness $D_z$ increases for a constant $d$
($\alpha$$\rightarrow$$\gamma$ or $\beta$$\rightarrow$$\delta$), the
magnetic field $B_z$ is seen to increase. This suggests that, by
depositing thicker micromagnets, larger magnetic fields can be
produced for the same separation between the micromagnet and the QW.
In Fig.~\ref{FigBz4rect}, we also show contour plots of $B_z$  on
the XY plane, which indicate the spatial extent of the magnetic
field and its gradient. In all cases ($\alpha$, $\beta$, $\gamma$
and $\delta$), we see that $B_z$ is confined to a narrow region in
the \mbox{x-direction}, and is delocalized in the \mbox{y-direction}
over the distance of two microns. With increasing thickness $D_z$
($\alpha$$\rightarrow$$\gamma$ or $\beta$$\rightarrow$$\delta$) the
"spread" of $B_z$ in the \mbox{x-direction} is seen to increase.

Our calculations show that at $d$=10~nm ($D_z$=150~nm), the maximum
value of the magnetic field is $|\vec{B}|_{max}$=0.6~T. This value
is larger than for a micro-disk because in the present case the
thickness of the rectangular micromagnet is larger. We must
emphasize that the gradient of the magnetic field \cite{Pulizzi} is
as large as \mbox{2~mT/\AA} for $d$=10~nm, so that a precise
determination of $|\vec{B}|_{max}$ or of the magnetic field profile
is challenging even in the simple case of a single-domain phase.

In Fig.~\ref{FigRectAbsVsd} we present the absorption spectrum at
three distances $d$ between the QW and the micromagnet: $d$=10, 20,
and 30~nm. In all three cases thickness of the micromagnet was kept
at a constant value of $D_z$=150~nm. As before, zero energy is
chosen at the main absorption peak. At $d$=10~nm, the binding energy
is 66~meV, while at $d$=60~nm it is smaller by a factor of 2. This
follows from the fact that the further the micromagnet is from the
QW, the smaller is the magnetic field at the QW position (see also
Fig.~\ref{FigBz4rect}). Non-diagonal transitions are relatively
strong both in the case of the micro-disk and of the rectangular
micromagnet\cite{Redlinski2}. The separation between the peaks
decreases as $d$ increases, because the gradient of the magnetic
field (and equivalently, the gradient of the potential) also
decreases with increasing $d$.

\begin{figure}[ht]
\includegraphics{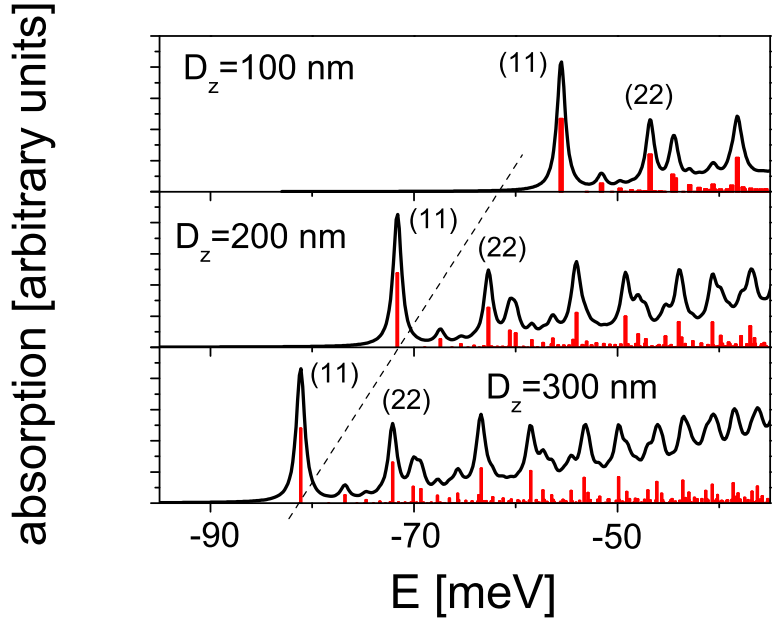}
\caption{Absorption coefficient for different thicknesses $D_z$ of a
rectangular micromagnet. In this case the distance between the
micromagnet and  the QW is constant, $d$=10~nm. Separation between
peaks is barely distinguishable for all three $D_z$ values. Note
that the absorption peak (11) shifts almost linearly with
$D_z$.}\label{FigRectAbsVsDz}
\end{figure}

\begin{figure}[ht]
\includegraphics{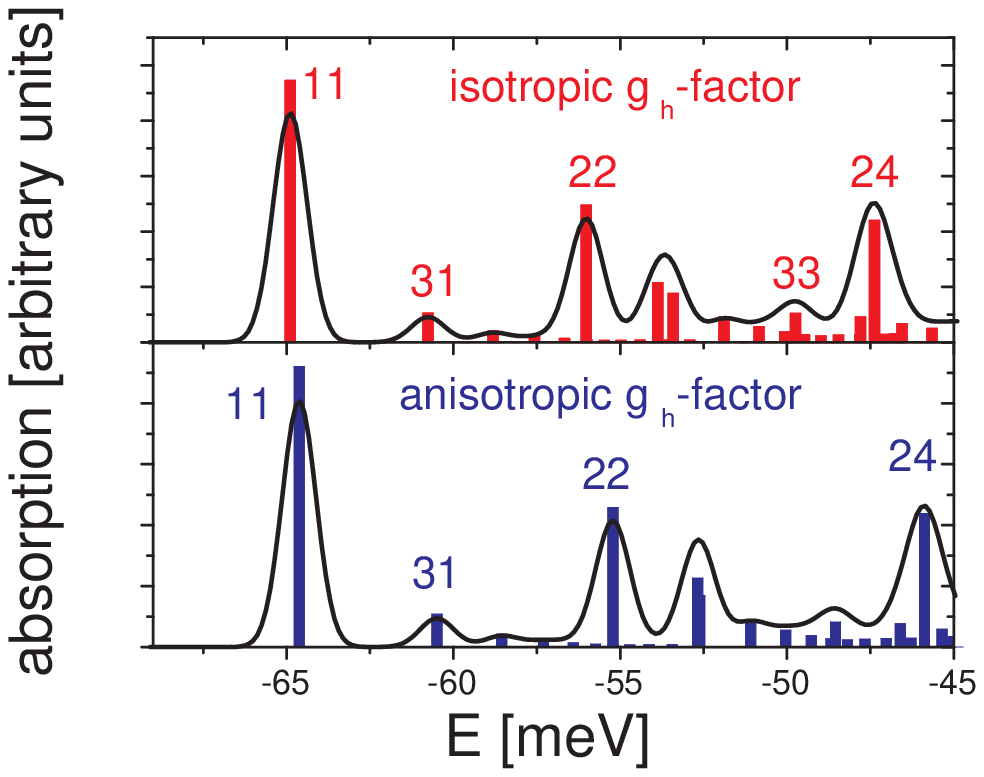}
\caption{Comparison of the absorption spectrum in two models: the
isotropic and the anisotropic g-factor of the hole. The electron
g-factor is isotropic. The top and bottom spectra are very similar.
Calculations done for rectangular Fe micromagnet with dimensions of
$D_x$=6$\mu$m, $D_y$=2$\mu$m, $D_z$=0.15$\mu$m on the top of a QW
structure at a distance $d$=10~nm apart.} \label{FigRectAbsAniso}
\end{figure}

In Fig.~\ref{FigRectAbsVsDz} we show the behavior of the absorption
coefficient for three thicknesses $D_z$=100, 200, and 300~nm, for
the same micromagnet-QW distance $d$.
With increasing $D_z$, the binding energy increases almost linearly.
This is seen in Fig.~\ref{FigRectAbsVsDz} as a linear shift of the
(11) transition. An interesting observation is that the separation
between the peaks does not depend on $D_z$ in the range 100-300~nm
and for the parameters which we have used. As in the previous
paragraphs the pair ($n_cn_v$) denotes transition between the
$n^{th}_v$ hole state and the $n^{th}_c$ electron state. Contrary to
the disk case, for the rectangular micro-magnet only (odd, odd) or
(even, even) transitions are allowed, see Fig~\ref{FigRectAbsAniso}.

Let us now return to the anisotropy of the g-factor of the holes
that was described in the theoretical part of this paper. In
Fig.~\ref{FigRectAbsAniso} we compare the absorption spectrum
calculated for the two approximations: the isotropic and the
anisotropic cases of the \mbox{$g_h$-factor}. As an example we have
chosen the rectangular micromagnet for discussing the magnetic
anisotropy of the QW.
The electron \mbox{g-factor} is isotropic in both cases. We see
that, at least in the low energy region, the two spectra are the
same: the position of the transition energies as well as the
oscillator strengths are nearly the same for both g-factor models.
Thus, for the arrangement discussed in this work (see inset in
Fig.~\ref{FigRectAbsVsd}) and for the geometrical parameters we have
used (such as the size and the shape of the magnets), only the
z-component of the magnetic field produced by the micromagnet is
important.

\section{Conclusions}
We analyzed theoretically the optical response of a hybrid structure
composed of a micromagnet deposited on top of a diluted magnetic
semiconductor quantum well structure. We specifically analyzed the
absorption coefficient which is directly related to the optical
absorption experiment, but the results of this paper are equally
well suited for photoluminescence experiment if only the ground
state is of interest.

The calculations were performed for two types of ferromagnetic
micromagnets: a cylindrical micro-disk and a rectangular
micromagnet. We analyzed the local magnetic field produced by the
micromagnets. In the case of a micro-disk this field, together with
the QW confinement potential, traps the quasi-particle (e.g., an
exciton) in all three spatial directions. However, a rectangular
micromagnet traps quasi-particles only in two spatial directions,
allowing it to behave as a one-dimensional quasi-particle. We
described the approximations we used in our approach, including a
discussion of the anisotropy of the \mbox{g-factor} of the hole
states. Then we calculated the absorption coefficients for both
shapes of the micromagnets for various micromagnet-QW separations
and micromagnet thicknesses.

In order to observe zero-dimensional and one-dimensional states
inside the DMS QW it is necessary to produce as strong a local
magnetic field as possible. This can be achieved, for example, by
utilizing materials with high saturation magnetization. Our analysis
shows that it is better to deposit a thicker ferromagnetic layer,
since thicker micromagnets produce a stronger local magnetic field.
As was expected, the growth of a high quality QW relatively close to
the ferromagnetic/semiconductor interface is of major importance for
optimal quasi-particle localization. Since high values of
\mbox{g-factors} are of critical importance for fabricating
efficient spin traps, optical localization is most likely expected
to be observed in DMS-based quantum structures at low temperature.

Finally, since quasi one-dimensional states emerge only below the
poles of the rectangular micromagnet and zero-dimensional states
emerge only below the center of the disk, spatially-resolved
techniques such as micro-photoluminescence, micro-reflectance or
near-field scanning optical microscopy are preferred for observing
the effects presented in this paper. Since both zero- and
one-dimensional states are spin polarized, for unambiguous
identification of the confined states the use of polarization
sensitive techniques are required. Specifically, in the case of a
rectangular micromagnet the states created underneath opposite poles
should have opposite spin-polarization which can serve as a
convincing test of the appearance of these new states.



\begin{thebibliography}{23}
\expandafter\ifx\csname
natexlab\endcsname\relax\def\natexlab#1{#1}\fi
\expandafter\ifx\csname bibnamefont\endcsname\relax
  \def\bibnamefont#1{#1}\fi
\expandafter\ifx\csname bibfnamefont\endcsname\relax
  \def\bibfnamefont#1{#1}\fi
\expandafter\ifx\csname citenamefont\endcsname\relax
  \def\citenamefont#1{#1}\fi
\expandafter\ifx\csname url\endcsname\relax
  \def\url#1{\texttt{#1}}\fi
\expandafter\ifx\csname urlprefix\endcsname\relax\def\urlprefix{URL
}\fi \providecommand{\bibinfo}[2]{#2}
\providecommand{\eprint}[2][]{\url{#2}}

\bibitem[{\citenamefont{Ohno et~al.}(2002)\citenamefont{Ohno, Matsukura, and
  Ohno}}]{Ohno1}
\bibinfo{author}{\bibfnamefont{H.}~\bibnamefont{Ohno}},
  \bibinfo{author}{\bibfnamefont{F.}~\bibnamefont{Matsukura}},
  \bibnamefont{and} \bibinfo{author}{\bibfnamefont{Y.}~\bibnamefont{Ohno}},
  \bibinfo{journal}{Japan Society of Applied Physics}
  \textbf{\bibinfo{volume}{5}}, \bibinfo{pages}{4} (\bibinfo{year}{2002}).

\bibitem[{\citenamefont{Grundler et~al.}(2004)\citenamefont{Grundler,
  Hengstmann, and Rolff}}]{Grundler}
\bibinfo{author}{\bibfnamefont{D.}~\bibnamefont{Grundler}},
  \bibinfo{author}{\bibfnamefont{T.~M.} \bibnamefont{Hengstmann}},
  \bibnamefont{and} \bibinfo{author}{\bibfnamefont{H.}~\bibnamefont{Rolff}},
  \bibinfo{journal}{Brazilian Journal of Physics}
  \textbf{\bibinfo{volume}{34}}, \bibinfo{pages}{598} (\bibinfo{year}{2004}).

\bibitem[{\citenamefont{Kreuzer et~al.}(2003)\citenamefont{Kreuzer, Rahm,
  Bigerger, Pulwey, Raabe, Schuh, Wegscheider, and Weiss}}]{Kreuzer}
\bibinfo{author}{\bibfnamefont{S.}~\bibnamefont{Kreuzer}},
  \bibinfo{author}{\bibfnamefont{M.}~\bibnamefont{Rahm}},
  \bibinfo{author}{\bibfnamefont{J.}~\bibnamefont{Bigerger}},
  \bibinfo{author}{\bibfnamefont{R.}~\bibnamefont{Pulwey}},
  \bibinfo{author}{\bibfnamefont{J.}~\bibnamefont{Raabe}},
  \bibinfo{author}{\bibfnamefont{D.}~\bibnamefont{Schuh}},
  \bibinfo{author}{\bibfnamefont{W.}~\bibnamefont{Wegscheider}},
  \bibnamefont{and} \bibinfo{author}{\bibfnamefont{D.}~\bibnamefont{Weiss}},
  \bibinfo{journal}{Physica E} \textbf{\bibinfo{volume}{16}},
  \bibinfo{pages}{137} (\bibinfo{year}{2003}).

\bibitem[{\citenamefont{Kossacki et~al.}(1999)\citenamefont{Kossacki, Cibert,
  Ferrand, Merle~d'Aubign\'{e}, Arnoult, Wasiela, Tatarenko, and
  Gaj}}]{Kossacki}
\bibinfo{author}{\bibfnamefont{P.}~\bibnamefont{Kossacki}},
  \bibinfo{author}{\bibfnamefont{J.}~\bibnamefont{Cibert}},
  \bibinfo{author}{\bibfnamefont{D.}~\bibnamefont{Ferrand}},
  \bibinfo{author}{\bibfnamefont{Y.}~\bibnamefont{Merle~d'Aubign\'{e}}},
  \bibinfo{author}{\bibfnamefont{A.}~\bibnamefont{Arnoult}},
  \bibinfo{author}{\bibfnamefont{A.}~\bibnamefont{Wasiela}},
  \bibinfo{author}{\bibfnamefont{S.}~\bibnamefont{Tatarenko}},
  \bibnamefont{and} \bibinfo{author}{\bibfnamefont{J.~A.} \bibnamefont{Gaj}},
  \bibinfo{journal}{Phys. Rev. B} \textbf{\bibinfo{volume}{60}},
  \bibinfo{pages}{16018} (\bibinfo{year}{1999}).

\bibitem[{\citenamefont{Combescot et~al.}(2004)\citenamefont{Combescot,
  Betbeder-Matibet, and Dubin}}]{Combescot}
\bibinfo{author}{\bibfnamefont{M.}~\bibnamefont{Combescot}},
  \bibinfo{author}{\bibfnamefont{O.}~\bibnamefont{Betbeder-Matibet}},
  \bibnamefont{and} \bibinfo{author}{\bibfnamefont{F.}~\bibnamefont{Dubin}},
  \bibinfo{journal}{European Physical Journal B} \textbf{\bibinfo{volume}{42}},
  \bibinfo{pages}{63} (\bibinfo{year}{2004}).

\bibitem[{\citenamefont{Freire et~al.}(2000)\citenamefont{Freire, Peeters,
  Matulis, Freire, and Farias}}]{Peeters1}
\bibinfo{author}{\bibfnamefont{J.~A.~K.} \bibnamefont{Freire}},
  \bibinfo{author}{\bibfnamefont{F.~M.} \bibnamefont{Peeters}},
  \bibinfo{author}{\bibfnamefont{A.}~\bibnamefont{Matulis}},
  \bibinfo{author}{\bibfnamefont{V.~N.} \bibnamefont{Freire}},
  \bibnamefont{and} \bibinfo{author}{\bibfnamefont{G.~A.}
  \bibnamefont{Farias}}, \bibinfo{journal}{Phys. Rev. B}
  \textbf{\bibinfo{volume}{62}}, \bibinfo{pages}{7316} (\bibinfo{year}{2000}).

\bibitem[{\citenamefont{Berciu and Jank\'{o}}(2003)}]{Berciu1}
\bibinfo{author}{\bibfnamefont{M.}~\bibnamefont{Berciu}} \bibnamefont{and}
  \bibinfo{author}{\bibfnamefont{B.}~\bibnamefont{Jank\'{o}}},
  \bibinfo{journal}{Phys. Rev. Lett.} \textbf{\bibinfo{volume}{90}},
  \bibinfo{pages}{246804} (\bibinfo{year}{2003}).

\bibitem[{\citenamefont{Kossut et~al.}(2001)\citenamefont{Kossut, Yamakawa,
  Nakamura, Cywi\'{n}ski, Fronc, Czeczott, Wr\'{o}bel, Kyrychenko, Wojtowicz,
  and Takeyama}}]{Kossut1}
\bibinfo{author}{\bibfnamefont{J.}~\bibnamefont{Kossut}},
  \bibinfo{author}{\bibfnamefont{I.}~\bibnamefont{Yamakawa}},
  \bibinfo{author}{\bibfnamefont{A.}~\bibnamefont{Nakamura}},
  \bibinfo{author}{\bibfnamefont{G.}~\bibnamefont{Cywi\'{n}ski}},
  \bibinfo{author}{\bibfnamefont{K.}~\bibnamefont{Fronc}},
  \bibinfo{author}{\bibfnamefont{M.}~\bibnamefont{Czeczott}},
  \bibinfo{author}{\bibfnamefont{J.}~\bibnamefont{Wr\'{o}bel}},
  \bibinfo{author}{\bibfnamefont{F.}~\bibnamefont{Kyrychenko}},
  \bibinfo{author}{\bibfnamefont{T.}~\bibnamefont{Wojtowicz}},
  \bibnamefont{and} \bibinfo{author}{\bibfnamefont{S.}~\bibnamefont{Takeyama}},
  \bibinfo{journal}{App. Phys. Lett.} \textbf{\bibinfo{volume}{79}},
  \bibinfo{pages}{1789} (\bibinfo{year}{2001}).

\bibitem[{\citenamefont{Cywi\'{n}ski et~al.}(2002)\citenamefont{Cywi\'{n}ski,
  Czeczott, Wr\'{o}bel, Fronc, Aleszkiewicz, Ma\'{c}kowski, Wojtowicz, and
  Kossut}}]{Cywinski}
\bibinfo{author}{\bibfnamefont{G.}~\bibnamefont{Cywi\'{n}ski}},
  \bibinfo{author}{\bibfnamefont{M.}~\bibnamefont{Czeczott}},
  \bibinfo{author}{\bibfnamefont{J.}~\bibnamefont{Wr\'{o}bel}},
  \bibinfo{author}{\bibfnamefont{K.}~\bibnamefont{Fronc}},
  \bibinfo{author}{\bibfnamefont{A.}~\bibnamefont{Aleszkiewicz}},
  \bibinfo{author}{\bibfnamefont{S.}~\bibnamefont{Ma\'{c}kowski}},
  \bibinfo{author}{\bibfnamefont{T.}~\bibnamefont{Wojtowicz}},
  \bibnamefont{and} \bibinfo{author}{\bibfnamefont{J.}~\bibnamefont{Kossut}},
  \bibinfo{journal}{Physica E} \textbf{\bibinfo{volume}{13}},
  \bibinfo{pages}{560} (\bibinfo{year}{2002}).

\bibitem[{\citenamefont{Redli\'{n}ski et~al.}(2004)\citenamefont{Redli\'{n}ski,
  Wojtowicz, Rappoport, Lib\'{a}l, Furdyna, and Jank\'{o}}}]{Redlinski2}
\bibinfo{author}{\bibfnamefont{P.}~\bibnamefont{Redli\'{n}ski}},
  \bibinfo{author}{\bibfnamefont{T.}~\bibnamefont{Wojtowicz}},
  \bibinfo{author}{\bibfnamefont{T.~G.} \bibnamefont{Rappoport}},
  \bibinfo{author}{\bibfnamefont{A.}~\bibnamefont{Lib\'{a}l}},
  \bibinfo{author}{\bibfnamefont{J.~K.} \bibnamefont{Furdyna}},
  \bibnamefont{and}
  \bibinfo{author}{\bibfnamefont{B.}~\bibnamefont{Jank\'{o}}},
  \bibinfo{journal}{submitted to Appl. Phys. Lett.}  (\bibinfo{year}{2004}).

\bibitem[{\citenamefont{Furdyna}(1988)}]{Furdyna2}
\bibinfo{author}{\bibfnamefont{J.~K.} \bibnamefont{Furdyna}},
  \bibinfo{journal}{J. Appl. Phys.} \textbf{\bibinfo{volume}{64}},
  \bibinfo{pages}{R29} (\bibinfo{year}{1988}).

\bibitem[{\citenamefont{Furdyna and Kossut}(1988)}]{Furdyna1}
\bibinfo{author}{\bibfnamefont{J.~K.} \bibnamefont{Furdyna}} \bibnamefont{and}
  \bibinfo{author}{\bibfnamefont{J.}~\bibnamefont{Kossut}},
  \emph{\bibinfo{title}{Diluted Magnetic Semiconductors}}
  (\bibinfo{publisher}{Academic}, \bibinfo{address}{Boston},
  \bibinfo{year}{1988}).

\bibitem[{\citenamefont{Dietl et~al.}(1991)\citenamefont{Dietl, Sawicki, Dahl,
  Heiman, Isaacs, Graf, Gubarev, and Alov}}]{Dietl1}
\bibinfo{author}{\bibfnamefont{T.}~\bibnamefont{Dietl}},
  \bibinfo{author}{\bibfnamefont{M.}~\bibnamefont{Sawicki}},
  \bibinfo{author}{\bibfnamefont{M.}~\bibnamefont{Dahl}},
  \bibinfo{author}{\bibfnamefont{D.}~\bibnamefont{Heiman}},
  \bibinfo{author}{\bibfnamefont{E.~D.} \bibnamefont{Isaacs}},
  \bibinfo{author}{\bibfnamefont{M.~J.} \bibnamefont{Graf}},
  \bibinfo{author}{\bibfnamefont{S.~I.} \bibnamefont{Gubarev}},
  \bibnamefont{and} \bibinfo{author}{\bibfnamefont{D.~L.} \bibnamefont{Alov}},
  \bibinfo{journal}{Phys. Rev. B} \textbf{\bibinfo{volume}{43}},
  \bibinfo{pages}{3154} (\bibinfo{year}{1991}).

\bibitem[{\citenamefont{Kuhn-Heinrich et~al.}(1993)\citenamefont{Kuhn-Heinrich,
  Ossau, Heinke, Fischer, Liz, Waag, and Landwehr}}]{Kuhn2}
\bibinfo{author}{\bibfnamefont{B.}~\bibnamefont{Kuhn-Heinrich}},
  \bibinfo{author}{\bibfnamefont{W.}~\bibnamefont{Ossau}},
  \bibinfo{author}{\bibfnamefont{H.}~\bibnamefont{Heinke}},
  \bibinfo{author}{\bibfnamefont{F.}~\bibnamefont{Fischer}},
  \bibinfo{author}{\bibfnamefont{T.}~\bibnamefont{Liz}},
  \bibinfo{author}{\bibfnamefont{A.}~\bibnamefont{Waag}}, \bibnamefont{and}
  \bibinfo{author}{\bibfnamefont{G.}~\bibnamefont{Landwehr}},
  \bibinfo{journal}{Applied Physics Letters} \textbf{\bibinfo{volume}{63}},
  \bibinfo{pages}{2932} (\bibinfo{year}{1993}).

\bibitem[{\citenamefont{Madelung}(1996)}]{Madelung}
\bibinfo{author}{\bibfnamefont{O.}~\bibnamefont{Madelung}},
  \emph{\bibinfo{title}{Introduction to Solid-State Theory}}
  (\bibinfo{publisher}{Springer, New York}, \bibinfo{year}{1996}),
  \bibinfo{edition}{3rd} ed.

\bibitem[{\citenamefont{Luttinger and Kohn}(1955)}]{Lutt}
\bibinfo{author}{\bibfnamefont{J.~M.} \bibnamefont{Luttinger}}
  \bibnamefont{and} \bibinfo{author}{\bibfnamefont{W.}~\bibnamefont{Kohn}},
  \bibinfo{journal}{Phys. Rev.} \textbf{\bibinfo{volume}{97}},
  \bibinfo{pages}{969} (\bibinfo{year}{1955}).

\bibitem[{\citenamefont{Abolfath et~al.}(2001)\citenamefont{Abolfath,
  Jungwirth, Brum, and MacDonald}}]{Abolfath}
\bibinfo{author}{\bibfnamefont{M.}~\bibnamefont{Abolfath}},
  \bibinfo{author}{\bibfnamefont{T.}~\bibnamefont{Jungwirth}},
  \bibinfo{author}{\bibfnamefont{J.}~\bibnamefont{Brum}}, \bibnamefont{and}
  \bibinfo{author}{\bibfnamefont{A.~H.} \bibnamefont{MacDonald}},
  \bibinfo{journal}{Phys. Rev. B} \textbf{\bibinfo{volume}{63}},
  \bibinfo{pages}{054418} (\bibinfo{year}{2001}).

\bibitem[{\citenamefont{Jackson}(1999)}]{Jackson}
\bibinfo{author}{\bibfnamefont{J.~D.} \bibnamefont{Jackson}},
  \emph{\bibinfo{title}{Classical electrodynamics}} (\bibinfo{publisher}{Wiley,
  New York}, \bibinfo{year}{1999}), \bibinfo{edition}{xxi} ed.

\bibitem[{\citenamefont{Kuhn-Heinrich et~al.}(1994)\citenamefont{Kuhn-Heinrich,
  Ossau, Bangert, Waag, and Landwehr}}]{Kuhn1}
\bibinfo{author}{\bibfnamefont{B.}~\bibnamefont{Kuhn-Heinrich}},
  \bibinfo{author}{\bibfnamefont{W.}~\bibnamefont{Ossau}},
  \bibinfo{author}{\bibfnamefont{E.}~\bibnamefont{Bangert}},
  \bibinfo{author}{\bibfnamefont{A.}~\bibnamefont{Waag}}, \bibnamefont{and}
  \bibinfo{author}{\bibfnamefont{G.}~\bibnamefont{Landwehr}},
  \bibinfo{journal}{Solid State Communications} \textbf{\bibinfo{volume}{91}},
  \bibinfo{pages}{413} (\bibinfo{year}{1994}).

\bibitem[{\citenamefont{Ha et~al.}(2003)\citenamefont{Ha, Hertel, and
  Kirschner}}]{Ha}
\bibinfo{author}{\bibfnamefont{J.~K.} \bibnamefont{Ha}},
  \bibinfo{author}{\bibfnamefont{R.}~\bibnamefont{Hertel}}, \bibnamefont{and}
  \bibinfo{author}{\bibfnamefont{J.}~\bibnamefont{Kirschner}},
  \bibinfo{journal}{Phys. Rev. B} \textbf{\bibinfo{volume}{67}},
  \bibinfo{pages}{224432} (\bibinfo{year}{2003}).

\bibitem[{\citenamefont{Crowell et~al.}(1997)\citenamefont{Crowell, Nikitin,
  Awschalom, Flack, Samarth, and Prinz}}]{Crowell}
\bibinfo{author}{\bibfnamefont{P.~A.} \bibnamefont{Crowell}},
  \bibinfo{author}{\bibfnamefont{V.}~\bibnamefont{Nikitin}},
  \bibinfo{author}{\bibfnamefont{D.~D.} \bibnamefont{Awschalom}},
  \bibinfo{author}{\bibfnamefont{F.}~\bibnamefont{Flack}},
  \bibinfo{author}{\bibfnamefont{N.}~\bibnamefont{Samarth}}, \bibnamefont{and}
  \bibinfo{author}{\bibfnamefont{G.~A.} \bibnamefont{Prinz}},
  \bibinfo{journal}{Journal of Applied Physics} \textbf{\bibinfo{volume}{81}},
  \bibinfo{pages}{5441} (\bibinfo{year}{1997}).

\bibitem[{\citenamefont{Donahue and Porter}(1999)}]{oommf}
\bibinfo{author}{\bibfnamefont{M.~J.} \bibnamefont{Donahue}} \bibnamefont{and}
  \bibinfo{author}{\bibfnamefont{D.~G.} \bibnamefont{Porter}},
  \bibinfo{journal}{Interagency Report NISTIR 6376, National Institute of
  Standards and Technology}  (\bibinfo{year}{1999}).

\bibitem[{\citenamefont{Pulizzi et~al.}(2000)\citenamefont{Pulizzi,
  Christianen, Maan, Wojtowicz, Karczewski, and Kossut}}]{Pulizzi}
\bibinfo{author}{\bibfnamefont{F.}~\bibnamefont{Pulizzi}},
  \bibinfo{author}{\bibfnamefont{P.~C.~M.} \bibnamefont{Christianen}},
  \bibinfo{author}{\bibfnamefont{J.~C.} \bibnamefont{Maan}},
  \bibinfo{author}{\bibfnamefont{T.}~\bibnamefont{Wojtowicz}},
  \bibinfo{author}{\bibfnamefont{G.}~\bibnamefont{Karczewski}},
  \bibnamefont{and} \bibinfo{author}{\bibfnamefont{J.}~\bibnamefont{Kossut}},
  \bibinfo{journal}{Phys. Stat. Sol a} \textbf{\bibinfo{volume}{178}},
  \bibinfo{pages}{33} (\bibinfo{year}{2000}).

\end{thebibliography}

\end{document}